\documentclass[journal=jctcce,manuscript=article,layout=onecolumn]{achemso}

\usepackage{dcolumn}

\usepackage[T1]{fontenc}
\usepackage{bbold}
\usepackage{amsmath}
\usepackage{amssymb}
\usepackage{graphicx}
\usepackage{subscript}
\usepackage[unicode=true,pdfusetitle,
bookmarks=true,bookmarksnumbered=false,bookmarksopen=false,
breaklinks=false,pdfborder={0 0 1},backref=false,colorlinks=false]
{hyperref}
\usepackage{xcolor}
\usepackage[title]{appendix}

\usepackage{bm}
\usepackage{physics}
\usepackage[version=4]{mhchem}
\usepackage{textcomp}
\usepackage{indentfirst}
\usepackage{listings}
\usepackage{svg}
\usepackage[export]{adjustbox}
\usepackage{lipsum} 
\usepackage{ulem} 

\newcommand{\editor}[2]{%
  \expandafter\newcommand\csname #1note\endcsname[1]{%
    \textcolor{#2}{(\textbf{#1:} \it ##1)}}%
  \expandafter\newcommand\csname #1\endcsname[1]{%
    \textcolor{#2}{##1}}%
  \expandafter\newcommand\csname #1cancel\endcsname[1]{%
    \textcolor{#2}{\sout{##1}}}%
  \expandafter\newcommand\csname #1change\endcsname[2]{%
    \textcolor{#2}{\sout{##1} ##2}}%
  \newenvironment{#1text}{\color{#2}}{\color{black}}
}

\editor{DS}{blue}
\editor{nsc}{orange}

\usepackage{cancel}

\newcommand{\be}{\begin{equation}}
\newcommand{\ee}{\end{equation}}
\newcommand{\bea}{\begin{eqnarray}}
\newcommand{\eea}{\end{eqnarray}}
\newcommand{\nn} {\nonumber}

\newcommand{\br}{{\bf r}}

\newcommand{\bR}{{\bf R}}

\newcommand{\bq}{{\bf q}}

\newcommand{\bk}{{\bf k}}
\newcommand{\bzero}{{\mathbf{0}}}

\def\QE{\textsc{Quantum ESPRESSO}}
\setkeys{acs}{doi = true}

\title{Efficient nonequilibrium electron dynamics from first-principles: leveraging Koopmans spectral functionals and Wannier localization}

\author{Giovanni Cistaro}%
\email{giovannicistaro@gmail.com}
\affiliation{Theory and Simulation of Materials (THEOS), \'Ecole Polytechnique F\'ed\'erale de Lausanne (EPFL), CH-1015, Lausanne, Switzerland}

\author{Miguel S\'a}
\affiliation{Departamento de F\'isica de la Materia Condensada, Universidad Aut\'{o}noma de Madrid, E-28049 Madrid, Spain}

\author{Davide Sangalli}
\affiliation{Istituto di Struttura della Materia (ISM - CNR),  and European Theoretical Spectroscopy Facilities (ETSF), Piazza Leonardo da Vinci 32, 20133 Milano, Italy}

\author{Antonio Pic\'on}
\affiliation{Instituto de Ciencia de Materiales de Madrid (ICMM-CSIC), E-28049 Madrid, Spain}

\author{Nicola Colonna}
\affiliation{PSI Center for Scientific Computing, Theory and Data, Paul Scherrer Institute, 5232 Villigen PSI, Switzerland}

\begin{document}

\begin{abstract}
We present an efficient first-principles approach for simulating the nonequilibrium electron dynamics in extended systems beyond the linear regime. The method combines Koopmans-compliant functionals, which provide an accurate quasiparticle band structures, with the real-time evolution of the electronic density matrix in a Wannier basis within the Hartree plus screened exchange (HSEX) approximation. 
The locality of the orbital basis enables physically motivated approximations that significantly reduce both the computational cost and memory requirements while preserving accuracy. The screened Coulomb interaction, the central ingredient of the HSEX self-energy, is computed efficiently using density-functional perturbation theory. We benchmark the approach in the linear regime against experimental spectra and reference Green's function calculations for systems featuring both weakly and strongly bound excitons. Moving to the nonlinear regime, we investigate high-harmonic generation (HHG) in silicon and lithium fluoride. While in silicon the HHG spectrum is largely governed by the quasiparticle band structure, in LiF, a material featuring strong excitonics effect, the harmonic emission is selectively enhanced at excitonic resonances, suggesting that HHG probes correlated electron-hole excitations rather than solely the quasiparticle band structure. The present framework enables fully \textit{ab-initio} simulations of excitonic effects in nonlinear optical spectra at a significantly reduced computational cost compared to real-time Green's function approaches, providing an efficient route to the study of ultrafast and strong-field phenomena in solids.
\end{abstract}

\maketitle

\section{Introduction}
Ultrafast experimental setups ranging from table-top optics to x-ray free-electron lasers (XFEL) have developed tremendously over the last years allowing not only for the experimental exploration of time-dependent dynamics with unprecedented temporal resolution reaching even the attosecond timescale \cite{inzani2025}, but also for the control of the macroscopic properties of quantum materials via ultrafast light-matter interaction~\cite{de_la_torre_colloquium_2021}. 
The interpretation of such experimental measurements often requires theoretical and computational support to unravel the complexity of the data involved and the connection between the macroscopic property and its microscopic origin. 

Fully {\it ab-initio} modeling of ultrafast processes and nonequilibrium (NEQ) electron dynamics requires an inherently real-time framework~\cite{caruso2026}. Unlike linear-response approaches, real-time methods naturally describe nonlinear light-matter interactions and ultrafast electron dynamics. They also provide access to strongly driven nonequilibrium phenomena induced by intense laser fields, such as high-harmonic generation (HHG).
Two {\it ab-initio} real-time approaches are well-established in the community; i) real-time time-dependent density functional theory (TDDFT) and ii) NEQ many-body perturbation theory (NEQ-MBPT). TDDFT is in principle an exact theory for the time evolution of the electronic charge density, still its accuracy heavily relies on the approximate time-dependent exchange-correlation potential; within commonly used local or semi-local adiabatic approximations, TDDFT usually performs well for small molecules~\cite{jacquemin_extensive_2009}, but it is known to fail in describing Rydberg states and charge-transfer excitation due to an incorrect description of the asymptotic long-range exchange~\cite{botti_long-range_2004, botti_time-dependent_2007}. For extended systems, electron energy loss and inelastic X-ray spectra at small and large momentum transfer are often well reproduced within standard TDDFT~\cite{Martin_Reining_Ceperley_2016}. Instead, TDDFT fails in predicting the optical ($\mathbf{q}=0$) spectra of insulating and semiconducting systems~\cite{Onida2002}. This failure originates in the incorrect asymptotic behavior of the exchange-correlation kernel (the derivative of the exchange-correlation potential with respect to the density) which leads to missing completely excitonic effects (i.e. those resulting from the electron-hole interactions) in the absorption spectra of non-metallic systems~\cite{Onida2002, botti_long-range_2004}. Hybrid functionals, which incorporate a fraction of Fock exchange, partially alleviate these shortcomings and yield improved results~\cite{paier_dielectric_2008,yang_simple_2015,chen_nonempirical_2018,tal_accurate_2020,ohad_optical_2023}. Their performance can be further enhanced by using dielectric-dependent schemes, in which the fraction of exact exchange is scaled by the dielectric constant of the system~\cite{shimazaki_band_2008,marques_density-based_2011,skone_self-consistent_2014,brawand_generalization_2016, chen_nonempirical_2018, tal_accurate_2020}, or by employing optimally tuned hybrid functionals that satisfy exact constraints~\cite{wing_band_2020, stein_fundamental_2010, refaely-abramson_quasiparticle_2012, refaely-abramson_solid-state_2015,ohad_optical_2023}. However, hybrid-functionals TDDFT approaches come at the cost of increased computational expense and of introducing some level of empiricism in the choice of the parameters controlling the exchange partitioning. TDDFT based on Hubbard functionals has also shown some promise~\cite{WeiKu2010PRB}.


A viable, although more expensive, alternative to TDDFT is represented by NEQ-MBPT. Exchange and correlation effects due to the electron-electron interaction are encoded in the self-energy, typically evaluated at the GW level of approximation~\cite{Hedin1965}. 
To the best of our knowledge, the full real-time GW (RT-GW) approach has been applied only to model systems~\cite{myohanen_kadanoff-baym_2009, puig_von_friesen_kadanoff-baym_2010, von_friesen_successes_2009, perfetto_nonequilibrium_2015, schuler_time-dependent_2016, schlunzen_achieving_2020}, while a fully {\it ab-initio} treatment relies on additional approximations, particularly on the use of a static-screening scheme, i.e. the so-called Hatree plus screened exchange (HSEX) approximation, sometimes also called adiabatic-GW~\cite{attaccalite_real-time_2011,  perfetto_pump-driven_2019, chan_giant_2021, sangalli_excitons_2021, dogadov2026, mocatti_nonequilibrium_2026,marek_linear_2025}. Within the HSEX approximation and in the linear regime, RT-HSEX reduces to the well known GW plus Bethe Salpeter equation~\cite{attaccalite_real-time_2011} (BSE), which has proven to be highly accurate in describing optical excitations in non-metallic systems~\cite{Onida2002,Martin_Reining_Ceperley_2016}. Extending this framework to the time domain allow addressing out-of-equilibrium phenomena and ultrafast dynamics of electrons, holes, and their bound states (excitons) as revealed e.g. in transient absorption spectroscopies and/or time-resolved angle-resolved photo-emission spectroscopy~\cite{sangalli_excitons_2021, chan_giant_2021, chan_giant_2023}. 
Even within these approximations the full \textit{ab-initio} treatment of the problem remains a formidable numerical task because of i) the difficulties to evaluate the time evolution of the HSEX self-energy when using single-particle Bloch eigenstates as a basis, ii) the complex convergence of the GW calculation which needs to consider a large number of empty-states, and iii) the difficulties in a proper description of the coupling with the external field.\\

We present here a novel, efficient, and fully {\it ab-initio} implementation of NEQ-MBPT within the HSEX approximation that overcomes the limitations described above. This is achieved by integrating Koopmans functionals \cite{linscott_koopmans_2023} with the \verb|EDUS| code \cite{cistaro2022theoretical}.
\verb|EDUS| was originally developed to evolve the density matrix in real-time using a Wannier basis \cite{PhysRevResearch.3.013144, malakhov2024exciton, quintela2026excitonic, mosquera2024topological}, which allows to avoid divergences related to the diabatic changes in the basis and to reduce the computational cost by exploiting the locality of Wannier function. In addition, the equilibrium quasiparticle electronic structure -- the starting point for the real-time dynamics -- is obtained using Koopmans-compliant functionals~\cite{dabo_koopmans_2010, borghi_koopmans-compliant_2014, colonna_jctc_2018, nguyen_koopmans_2018} , which provide GW-level spectral accuracy~\cite{schubert_testing_2023,nguyen_first-principles_2015,nguyen_first-principles_2016,elliott_koopmans_2019,colonna_jctc_2019, borghi_koopmans-compliant_2014, nguyen_koopmans_2018,de_almeida_electronic_2021,degennaro_bloch_2022,colonna_jctc_2022,ingall_accurate_2024, marrazzo_spin-dependent_2024, schubert_predicting_2024, stojkovic_predicting_2026} at significantly reduced computational cost and complexity~\cite{ferretti_bridging_2014, colonna_jctc_2019}. Finally, the calculation of the matrix elements of the HSEX self-energy is streamelined using the machinery of density functional perturbation theory (DFPT), thus avoiding explicit reference to empty-state.~\cite{baroni_phonons_2001} 
The resulting \verb|KCW|-\verb|EDUS| framework enables predictive first-principles simulations of ultrafast and nonlinear phenomena with an accuracy comparable to state-of-the-art RT-HSEX methods, while substantially reducing the computational cost.

As a validation of the implementation we present results for the optical absorption in the linear-regime of standard semiconductors and wide band-gap insulators and compare the results with experiments and state-of-the-art GW-BSE calculations.  Moving beyond linear response, we showcase the capabilities of the present implementation by investigating the effect of electron-electron interaction on the high-harmonic generation in bulk silicon and lithium fluoride when driven by short, intense near- and mid-infrared laser pulses, and highlight the importance of excitonic effects absent in the widely used independent particle approximation.

The paper is organized as follows. We first present the theoretical framework, including non-equilibrium Green's functions, Koopmans-compliant functionals, and Wannier functions. We then describe the first-principles procedure used to compute the bare and screened Coulomb matrix elements in the Wannier basis. Finally, we present our benchmark results, both in linear and non-linear regime.

\section{Theoretical framework} 
We outline here the theoretical framework of the present work, combining non-equilibrium Green's functions, Wannier interpolation, and Koopmans-compliant functionals to obtain a fully first-principles description of electron dynamics beyond the independent-particle approximation.

\subsection{Non-equilibrium Green's functions for electrons dynamics}
The time evolution of an electronic system under the action of an external field can, in principle, be fully described by the Kadanoff-Baym equations, i.e., the equations of motion for the nonequilibrium Green's function \cite{kadanoff1962quantum}. Their solution requires knowledge of the electron self-energy, which accounts for electron-electron interactions and is itself a functional of the one-particle Green's function.
Under the assumption that the self-energy is local in time and under the Generalized Kadanoff-Baym ansatz \cite{lipavsky1986generalized}, the time evolution for the non-equilibrium Green's function $G^<(t,t')$ can be recast into an equation of motion (EOM) for the one-body density matrix $\rho_{nm}({\bf k},t)$:
\begin{equation}
    \rho_{nm}(\textbf{k},t) = \langle c^{\dagger}_{m\bf k}(t) c_{n\textbf{k}}(t)\rangle =  iG^{<}_{nm\bf k}(t,t), 
\end{equation}
where $c^\dagger_{n\bf k}(t)$ and $c_{n\bf k}(t)$ are respectively the creation and annihilation operators for an electron with quantum state described by $n \bf k$. The quantum number ${\bf k}$ corresponds to the quasi-momentum, while $n$ and $m$ identify the elements of the basis, in our particular case, they correspond to a set of Wannier orbitals that are used to construct the Bloch basis \cite{cistaro2022theoretical}. The EOM is then written as 
\begin{multline}\label{eq:EOM}
    i{{\partial \rho_{nm}(\textbf{k},t)}\over{\partial t}} = 
    [H(\textbf{k}) + \Delta\Sigma[\rho](\textbf{k},t), \rho(\textbf{k},t)]_{nm} \\ 
    +[\boldsymbol{\varepsilon}(t)\cdot\boldsymbol{\xi}(\textbf{k}), \rho(\textbf{k},t)]_{nm} + 
    i  \boldsymbol{\varepsilon}(t)\cdot\nabla_{\textbf{k}} \rho_{nm}(\textbf{k},t).
\end{multline}
In the expression above, $[\cdots,\cdots]$ indicates a commutator, $H$ is the single-particle Hamiltonian of the system encoding the equilibrium band structure of the material, and $\Delta\Sigma$ denotes the deviation from equilibrium of the self-energy encoding many-body interactions among electrons beyond ground-state.
$\boldsymbol{\varepsilon}(t)$ is the external time-dependent electric field, which couples with the Berry connection $\boldsymbol{\xi}$, and with the grandient of the density matrix.
In the present work we will consider only static self-energy with linear dependency on the density matrix. This means that $\Delta \Sigma[\rho]=\Sigma[\Delta\rho]$, where $\Delta \rho$ represents the time-dependent deviation of the density matrix from its equilibrium value (i.e., the density matrix before an external field is applied to the system). At equilibrium, $\Delta \rho = 0$, and consequently $\Sigma = 0$, with many-body effects already included in the effective quasi-particle hamiltonian $H$. When the system is driven out of equilibrium, $\Delta \rho \neq 0$, and $\Sigma$ becomes finite, capturing the electron--electron correlations within the excited (non-equilibrium) component of the system. \\
Several well-established expressions for the self-energy exist, each corresponding to a different level of treatment of electron-electron correlations. In the following, we present some of the most commonly used forms:
\begin{itemize}
    \item IPA (Independent Particles Approximation). No self-energy, electrons are fully independent and non-interacting. 
    \begin{equation}
        \Sigma^{\text{IPA}}_{nn'}[\Delta \rho](\textbf{k},t) = 0 \nonumber
    \end{equation}
    \item RPA (Random Phase Approximation). Each electron experiences a self-consistent, semiclassical Hartree potential generated by the charge density of all electrons. Exchange and correlation effects beyond this mean-field description are neglected.
    \begin{equation}
        \Sigma^{\rm{RPA}}[\Delta \rho](\textbf{k},t) = \Sigma^\text{H}[\Delta \rho](\textbf{k},t) \nonumber
    \end{equation}
    where
    \begin{equation}
    \label{eq:RPA_selfenergy}
       \Sigma^\text{H}[\rho]_{nn'}(\textbf{k},t) = \sum_{\alpha \beta \bf k'}  V_{{n{\bf k}\alpha {\bf k}', n'{\bf k}\beta\bf k'}}\rho_{\beta \alpha}({\bf k}') 
    \end{equation}
    \item HSEX (Hartree plus Screened EXchange). Electrons interact with the classical Hartree potential and with a Fock (exchange) potential, screened by the presence of the other electrons in the solid
    \begin{equation}
          \Sigma^{\text{HSEX}} =\Sigma^{\text{H}} + \Sigma^{\text{SEX}}\nonumber
    \end{equation}
    where 
    \begin{equation}
        \Sigma^{\text{SEX}}_{nn'}[\rho](\textbf{k},t) = -\sum_{\alpha \beta \bf k'}  W_{n{\bf k}\alpha {\bf k}', \beta{\bf k}'n'{\bf k}}\rho_{\beta \alpha}({\bf k}').
        \label{eq:HSEX_selfenergy}
    \end{equation}
\end{itemize}
In the expression above $V$ and $W$ represent the the bare and statically screened Coulomb interactions. The screened interaction is in general a non linear functional of the density matrix. Here we assume that equilibrium screening can be used through the time-dependent propagation, as commonly done in RT-HSEX schemes.~\cite{attaccalite_real-time_2011} The convention for the matrix elements is: 
\begin{eqnarray}
        O_{nmn'm' } := \int d^3r\, d^3 r' \, \psi_{n}^*({\bf r}) \psi_{m}^*({\bf r}')  \nonumber \\  \times O(\br,\br') \psi_{n'}({\bf r}) \psi_{m'}({\bf r}') \nonumber \\
        = \langle \psi_n \psi_m|\hat{O}|\psi_{n'}\psi_{m'} \rangle
\end{eqnarray}


The time evolution of the density matrix provides access to the expectation value of any one-body operator over the ensemble via: \begin{equation}
\langle O(t)\rangle = \text{Tr} (\hat{O}\rho(t))
\end{equation}
In particular, the expectation value of the velocity operator can be computed using the expression of its matrix elements in crystal-momentum representation: 
\begin{equation}
    {\bf v}_{nm}(\textbf{k}) = \nabla_{\bf k} H_{nm}(\textbf{k}) -i [H(\textbf{k}), \boldsymbol{\xi}(\textbf{k})]_{nm}
\end{equation}

The expectation value of the velocity operator directly yields the current density:
\begin{equation}
    {\bf J}(t) = -|e| \langle {\bf v}(t)\rangle
\end{equation}
and the complex optical conductivity is computed as the ratio between the current and the external applied electric field $E$:
\begin{equation}
    \sigma_{ij}(\omega) = \frac{J_i(\omega)}{ E_j(\omega)} 
\end{equation}

Furthermore, integrating in time the expectation value of the velocity gives the expectation value of the position operator $\langle {\bf r}(t)\rangle=\int_{-\infty}^t \langle {\bf v}(t')\rangle dt'$ and the polarization: 
\begin{equation}
    {\bf P}(t) = -|e| \langle {\bf r}(t)\rangle. 
\end{equation}
The absorption spectrum of a material is proportional to the imaginary part of the optical susceptibility~\cite{cistaro2022theoretical}:
\begin{equation}
    \chi_{ij}(\omega) = \frac{P_i(\omega)}{E_j(\omega)}
\end{equation}

Beyond the linear regime, the relevant observable for strong-field processes is the emitted radiation rather than the susceptibility. Differentiating the expectation value of the velocity operator yields the acceleration $\langle \mathbf{a}(t)\rangle = d \langle \mathbf{v}(t) \rangle/dt$, the radiated power follows via the Larmor formula:
\begin{equation} \label{HHG}
    P_{\text{rad}} = \frac{e^2 |\langle {\bf a}(\omega)\rangle|^2 }{6\pi\epsilon_0 c^3}
\end{equation}

This formula enables us to compute the microscopic HHG spectrum of the quantum system.

\subsection{Wannier basis and interpolation}
In this section, we highlight some of the advantages of using a Wannier basis. The EOM, given in Eq.~\eqref{eq:EOM}, can be expressed in any Bloch basis. However, its practical implementation typically requires representing all relevant quantities in such a basis. 
While Bloch eigenstates are the most natural choice of basis, they are often not the most convenient for evaluating many material properties, as they do not necessarily form a smooth gauge throughout the Brillouin zone. As a result, quantities involving derivatives with respect to crystal momentum, such as $\nabla_{\mathbf{k}} \psi_{n\mathbf{k}}(\mathbf{r})$ and, consequently, $\nabla_{\mathbf{k}} \rho_{nm\mathbf{k}}$, may be ill-defined in some regions of the Brillouin zone because $\psi_{n\mathbf{k}}(\mathbf{r})$ can exhibit discontinuities \cite{blount1962formalisms}.
A well-known solution \cite{marzari_maximally_2012}, that we will pursue in this work, is to adopt a more suitable basis, obtained by applying a unitary rotation to the Bloch eigenstates:
\begin{equation}
    \tilde{\psi}_{n\bf k}(\textbf{r}) = \sum_{m} U_{mn}(\textbf{k}) \psi_{m\textbf{k}}(\textbf{r}),
\end{equation}
where $U(\textbf{k})$ is an arbitrary unitary matrix that can be chosen~\cite{marzari_maximally_2012} such that the resulting Wannier functions
\begin{equation}
    \omega_{n\bf R} (\textbf{r}) = \frac{1}{N_{\bf k}} \sum_{\bf k} e^{-i\bf k \cdot \bf R} \tilde{\psi}_{n\bf k}(\textbf{r})
\end{equation}
are maximally localized, i.e., they minimize the spread functional
\begin{equation}
    \Omega = \sum_n \Big( \langle \omega_{n\bf 0}| r^2 |\omega_{n\bf 0}\rangle - |\langle \omega_{n\bf 0}| {\bf r} |\omega_{n\bf 0}\rangle|^2 \Big).
\end{equation}
Wannier functions are guaranteed to decay exponentially in gapped materials (semiconductors and insulators) \cite{brouder2007exponential, nenciu1983existence, des1964analytical} and polynomially in metals \cite{cornean_localised_2019, he2001exponential}.
\\
Other possible solutions to the gradient discontinuities involve the used of covariant derivatives~\cite{attaccalite_nonlinear_2013}, or local gauge fixing of the bloch wave-functions~\cite{chan_giant_2021}.
The use of Wannier function, however, also leads to other advantages. An 
essential one for the present work is the possibility of performing Wannier interpolation~\cite{marzari_maximally_2012}. For one-body operators, the matrix elements can be computed over a sparse $\bf R$ grid and then interpolated over a much denser $\bf k$-grid by exploiting the spatial decay of the Wannier functions:
\begin{equation} \label{OBOWannierR}
    O_{nm}(\textbf{k}) = \sum_{\textbf{R}} e^{i\bf k \cdot \bf R} O_{nm}(\textbf{R}),
\end{equation}
with
\begin{equation} \label{ElementOR}
    O_{nm}(\textbf{R}) = \langle \omega_{n\textbf{0}}| O |\omega_{m\textbf{R}}\rangle,
\end{equation}
where $|\omega_{m\mathbf{R}}\rangle$ denotes the $m$-th Wannier orbital localized in the unit cell at lattice vector $\mathbf{R}$. Note that Eq.~\eqref{OBOWannierR} exhibits a smooth dependence on $\mathbf{k}$, provided that the Wannier orbitals decay exponentially. This follows from the fact that the matrix elements defined in Eq.~\eqref{ElementOR} rapidly vanish for large $\mathbf{R}$.
Another key advantage, as we will discuss later, is the possibility to introduce further approximations, which significantly reduce the computational cost.

\subsection{Koopmans compliant functionals}
In this paragraph, we briefly review the theory of KC functionals, used in this work as an efficient and reliable alternative to expensive and highly parametrized GW calculations. For a complete and exhaustive description of the theory we also refer the reader to Ref.~\cite{linscott_koopmans_2023} and references therein. 

KC functionals enforce a generalized piece-wise linearity (PWL) condition of the total energy when adding/removing an electron to/from the system. When screening and relaxation effects are correctly accounted for~\cite{colonna_jctc_2018, nguyen_koopmans_2018}, this is akin of achieving a correct description of charged excitations, as revealed, e.g., in photoemission experiments. The generalized PWL condition is imposed adding to standard local or semi-local density functional an orbital density dependent correction $\Pi_i^{\rm KC}$, while the screening and relaxation effects are taken into account by orbital-dependent screening coefficients $\alpha_i$, so that the KC total energy functional reads:
\begin{equation}\label{KC::1}
    E^{\text{KC}}[\rho, \rho_i] = E^{\text{DFT}}[\rho] + \sum_{i} \alpha_{i} \Pi^{\rm{KC}}_{i}[\rho, \rho_i]
\end{equation}
where $\rho(\mathbf{r})= \sum_i \rho_i(\mathbf{r})$ is the total charge density and $\rho_i(\mathbf{r})=f_i|\phi_i(\mathbf{r})|^2$ is the orbital density associated to orbital $\phi_i(\mathbf{r})$ at occupation $f_i$. In this work we used the so-called quadratic Koopmans Integral (qKI) correction~\cite{colonna_jctc_2018, colonna_jctc_2022}:
\begin{align}
    \Pi_i^{\rm qKI} & = \frac{1}{2}f_i(1-f_i)\int d\br d\br ' \rho_i(\br) f_{\rm Hxc} (\br, \br') \rho_i(\br') 
\end{align}
where $f_{\text{Hxc}}$ is the Hartree plus exchange and correlation kernel, that is the second derivative of the underlying DFT functional with respect to the density.
Finally, the Koopmans compliance is imposed on the variational orbitals that  minimize the KC energy functional. Variational orbitals are typically localized in space and can be very well approximated by maximally-localized WFs (MLWFs)~\cite{marzari_maximally_2012,nguyen_koopmans_2018,colonna_jctc_2022}. 
This Koopmans-Wannier formulation~\cite{colonna_jctc_2022, marrazzo_spin-dependent_2024} is deployed as a one-shot correction to DFT eigenvalues and eigenvectors in the \verb|KCW| module of \QE. Remarkably, the KC approach maintains a simple functional formulation while being typically as accurate as state of the art in Green's function theory~\cite{nguyen_koopmans_2018,colonna_jctc_2019,colonna_jctc_2022,ingall_accurate_2024}, at a reduced computational cost and complexity. 


\section{First-principles implementation}

\label{sec:3}
Building on the theoretical framework introduced in the previous section, we now describe its practical implementation. We focus in particular on the approximations that exploit the locality of Wannier functions and on the DFPT-based evaluation of the Coulomb interaction, which together constitute the core methodological advances of the present work.
The implementation is based on the \verb|EDUS| code~\cite{cistaro2022theoretical}, which propagates the equation of motion (EOM) for the density matrix in a Wannier basis. To achieve a fully \textit{ab-initio} framework, \verb|EDUS| has been interfaced with \QE~\cite{giannozzi_advanced_2017, giannozzi_quantum_2009} and Wannier90~\cite{pizzi_wannier90_2020}, so that every quantity entering the EOM is computed from first principles. Specifically:
\begin{itemize}
    \item The quasi-particle Hamiltonian $H$ is obtained from a Koopmans spectral functional calculation of the band structures as implemented in the \verb|KCW| package~\cite{colonna_jctc_2022} of \QE, followed by a Wannierization. This provides an accurate description of the energy band gap and band structure at equilibrium. 
    \item The Berry connection $\boldsymbol{\xi}$ is obtained from the Wannierization, as the Fourier transform of the matrix elements of the position operator \cite{blount1962formalisms}, obtained as an output of the Wannier90 code~\cite{pizzi_wannier90_2020}.
    \item The self-energy is calculated using the matrix elements of the bare and screened Coulomb interaction over the Wannier functions using density functional perturbation theory and an extension of the \verb|KCW| module \cite{colonna_jctc_2022} of \QE\, as detailed below.
\end{itemize}

\subsection{The density-density approximation}
The locality and orthogonality of Wannier functions provides a natural route to simplifying the structure of the Coulomb interaction. In this section, we introduce the density-density approximation: we only retain the dominant contributions to the electron-electron interaction while substantially reducing the computational cost and memory requirements.

In the Wannier basis, the matrix elements of the two-body Coulomb interaction operators -- bare or screened -- are labeled by four independent band indices and three independent lattice vectors, one of which is fixed by translational symmetry:
\begin{align}
    &\mathcal{U}_{n\textbf{0}m\textbf{R}, n'\textbf{R}' m'\textbf{R}''} = \nonumber \\
    &\;\; \int d^3 r\, d^3 r' \; \omega^*_{n\textbf{0}}(\textbf{r}) \omega^*_{m\textbf{R}}(\textbf{r}')\,  \mathcal{U}(\textbf{r},\textbf{r}') \, \omega_{n'\textbf{R}'}(\textbf{r})\omega_{m'\textbf{R}''}(\textbf{r}') 
    \label{KC::3}
\end{align}
where $\mathcal{U}$ represents either the bare ($v$) or the screened ($w=\epsilon^{-1}v$) Coulomb kernel. Evaluating and storing this full object is, in general, extremely expensive for extended systems described by many Wannier functions.
Wannier functions are, however, well localized in real space, and by construction the overlap between orbitals is generally small. This locality has a direct consequence on the structure of
Eq.~\eqref{KC::3}. The ``diagonal'' contributions, built from orbital densities,
\begin{equation}
\rho_{n\mathbf R}(\mathbf r)=|\omega_n(\mathbf r-\mathbf R)|^2
\end{equation}
are positive-definite quantities. The ``off-diagonal'' contributions, instead, built from orbital coherences:
\begin{equation}
\rho_{nm\mathbf R}(\mathbf r)=\omega_n^*(\mathbf r)\omega_m(\mathbf r-\mathbf R).
\end{equation}
have many nodes within the support of the functions. As a consequence, $\rho_{nm\bf R}$ oscillates in sign precisely in the region where the Coulomb kernel is largest. Upon integration against $\mathcal{U}(\mathbf{r},\mathbf{r}')$, this oscillatory behavior produces substantial internal cancellation, so that coherence terms are expected to be much smaller in magnitude than the corresponding density terms. Moreover, when two Wannier orbitals have different centers, the overlap of the domain of two different Wannier functions is small, giving another valid reason to neglect the integrals built from the coherences. 
This observation motivates the density-density approximation: we retain only the contributions to Eq.~\eqref{KC::3} built from orbital densities -- i.e., $n'=n$, $m'=m$, $\mathbf{R}'=\mathbf{0}$, $\mathbf{R}''=\mathbf{R}$ -- and discard all terms originating from orbital coherences. The resulting quantity depends on only two band indices and a single lattice vector:
\begin{align}
    & \mathcal{U}_{nm}(\bR) := \mathcal{U}_{n{\bf 0} m\textbf{R}, n{\bf 0}m \textbf{R}} =  \nn \\ 
     &\quad\quad  = \int d^3 r\, d^3 r' \; |\omega_{n\textbf{R}}(\textbf{r})|^2\,  \mathcal{U}(\textbf{r},\textbf{r}') \, |\omega_{m\textbf{0}}(\textbf{r}')|^2 \nn \\
     & \quad\quad  =  \langle w_{n\bR} w_{m\bzero} | \hat{\mathcal{U}} | w_{n\bR} w_{m\bzero}\rangle
    \label{KC::4}
\end{align}
The density-density ansatz of Eq.~\eqref{KC::4}  can be derived as an exact condition if the Coulomb kernel is assumed slowly varying on the scale of the Wannier functions' support. In this limit, the kernel $\mathcal{U}(\mathbf{r},\mathbf{r}')$ can be taken out of the integration, and the four-orbital integral of Eq.~\eqref{KC::3} factorizes into a product of two overlap integrals. As a consequence, orthonormality of the Wannier functions enforces exactly $n'=n$, $m'=m$, $\mathbf{R}'=\mathbf{0}$, $\mathbf{R}''=\mathbf{R}$:
\begin{align}
    & \mathcal{U}_{n\textbf{0}m\textbf{R}, n'\textbf{R}' m'\textbf{R}''} \approx
    \mathcal{U}_{nm}(\mathbf{R})
    \nonumber \\
    & \ \ \ \ \times \left(\int d^3 r \, \omega^*_{n\bf 0}(\textbf{r})\omega_{n'\textbf{R}'}(\textbf{r}) \right) \nonumber \\ 
    & \ \ \ \  \ \ \ \ \times \left(\int d^3 r' \omega^*_{m\bf R}(\textbf{r}')\omega_{m'\textbf{R}''}(\textbf{r}')\right)\nonumber
    \\
    &= \mathcal{U}_{nm}(\mathbf{R})\delta_{nn'}\delta_{\bf R' 0} \delta_{mm'}\delta_{\bf R R''}
\end{align}

Recasting Eq.~\eqref{KC::4} in reciprocal space offers a complementary view of the same approximation:
\begin{equation}
    \mathcal{U}_{n'\textbf{k}'-\textbf{q}m'\textbf{k}+  \textbf{q}, n\textbf{k}'m\textbf{k}} = \frac{\delta_{n'n}\delta_{m'm}}{N_k^3} \sum_{\textbf{R}} e^{i\bf q \cdot R} \mathcal{U}_{nm}(\textbf{R})
\end{equation}
Only the dependence on the momentum transfer $\mathbf{q}$ survives, while the dependence on the individual momenta $\mathbf{k}$ and $\mathbf{k}'$ is dropped. 
The Coulomb interactions defining the self-energies in Eqs. ~\eqref{eq:RPA_selfenergy} and ~\eqref{eq:HSEX_selfenergy}, are recovered from the equation above for a specific choice of the transfer momentum $\bq$, that is $\bf q=0$, and $\bf q=k'-k$ for the Hartree and SEX self-energy, respectively. 
This simplification is tied to the Wannier gauge: rotated back to the basis of Bloch eigenstates, the same matrix would in general retain a full dependence on both $\mathbf{k}$ and $\mathbf{k}'$. The density-density approximation is therefore best understood -- and is only justified -- in a sufficiently localized basis.
This general considerations and physical justifications are further supported by the validation results presented in Sec.~\ref{sec:Linear_regime}.

Beyond its physical motivation, the approximation carries a substantial practical benefit: it reduces the four-index Coulomb vertex of Eq.~\eqref{KC::3}, which formally contains $N_w^4 N_R^3$ independent elements, to a two-index matrix for each lattice vector $\mathbf{R}$, i.e., only $N_w^2 N_R$ independent quantities. Here $N_w$ is the number of Wannier functions, while $N_R$ the number of lattice vectors $\mathbf{R}$, or equivalently of $\bf k$ vectors, used in the simulations. For systems described by many Wannier functions, this reduction translates directly into the memory and computational savings quantified in
Sec.~\ref{sec:HSEX_dd}.  

\subsection{Self-energies within the density-density approximation}
\label{sec:HSEX_dd}
One of the main advantages of the density-density approximation is the simplification of  Eqs.~\eqref{eq:RPA_selfenergy} and \eqref{eq:HSEX_selfenergy}, which significantly reduces the computational cost associated with the evaluation of both $\Sigma^\text{H}$ and $\Sigma^\text{SEX}$. The results obtained in this paragraph are summarized in Table \ref{tab:hartree_sex_scaling} together with the memory needed to store the matrix elements of bare and screened Coulomb interaction. \\
Within this ansatz, the self-energies can be expressed as the Fourier transform of the following quantities \cite{molinero2025semiconductorwannierequationsrealtime}:
\begin{eqnarray}
\label{SigmaH}
\Sigma^\text{H}_{nm}(\mathbf{R}) 
&=& \delta_{nm}\delta_{\mathbf{R},0}
\sum_{n'} \rho_{n'n'}(\mathbf{0}) 
\sum_{\mathbf{R}'} V_{nn'}(\mathbf{R}')
\\
\label{SigmaSEX}
\Sigma^{\text{SEX}}_{nm}(\mathbf{R}) 
&=& - W_{nm}(\mathbf{R}) \rho_{nm}(\mathbf{R}) .
\end{eqnarray}
We note that within the density-density approximation the Hartree term becomes diagonal in the Wannier basis. Equation~\eqref{SigmaH} can be easily interpreted as a  semiclassical energy, as no coherences appear.
The Coulomb potential acting on a given orbital $n$ is  due to the total charge density of all other orbitals $n'$ weighted by the matrix elements of the interaction $V_{nn'}(\mathbf{R'})$. The sum over $\mathbf{R}'$ reflects the fact that the interaction is with all of the Wannier functions of type $n'$ on any unit cell.
Moreover, the $\delta_{\mathbf{R,0}}$ in eq.~\eqref{SigmaH} implies that the Hartree self-energy remains diagonal also in the $\mathbf{R}$ space, i.e. it corrects only the charge terms and not the coherences.\\
We now focus on the computational cost of each expression we wrote to define the self-energy.  For Eq.~\eqref{SigmaH}, one needs to evaluate $N_w$ elements, each involving a sum over $N_w$ orbitals (we neglect the sum over $\mathbf{R}'$, since it can be precomputed once at the beginning of the calculation). The resulting computational cost scales as $\mathcal{O}(N_w^2)$.\\
For Eq.~\eqref{SigmaSEX}, instead, there are $N_w^2 N_R$ matrix elements that must be computed, but each evaluation contains only a simple multiplication of numbers. The resulting cost therefore scales as $\mathcal{O}(N_w^2 N_R)$.\\
By contrast, for the full expressions in Eqs.~\eqref{eq:RPA_selfenergy} and ~\eqref{eq:HSEX_selfenergy}, there are $N_w^2 N_R$ elements to be computed, and each of those requires a summation over $N_w^2 N_R$ contributions, leading to an overall scaling of $\mathcal{O}(N_w^4 N_R^2)$ for the evaluation of each self-energy.\\

\begin{table*}[h]
\centering
\begin{tabular}{|l|c|c|}
\hline
                          & \textbf{Computational Cost} & \textbf{Data Storage} \\ \hline
\textbf{Hartree (density-density)} & $\mathcal{O}(N_w^2N_R^0)$ & $N_w^2 N_R$ \\ \hline
\textbf{Hartree (full theory)}     & $\mathcal{O}(N_w^4 N_R^2)$ & $N_w^4 N_R^3$ \\ \hline
\textbf{SEX (density-density)}     & $\mathcal{O}(N_w^2 N_R)$ & $N_w^2 N_R$ \\ \hline
\textbf{SEX (full theory)}         & $\mathcal{O}(N_w^4 N_R^2)$ & $N_w^4 N_R^3$ \\ \hline

\end{tabular}
\caption{Scaling of computational cost and data storage for Hartree and SEX methods in density-density and full theory formulations.}
\label{tab:hartree_sex_scaling}
\end{table*}

\subsection{Coulomb matrix elements from DFPT}
Having established the simplified form of the self-energies within the density-density approximation, we now turn to the evaluation of the corresponding bare and screened Coulomb matrix elements. These quantities can be computed efficiently within DFPT~\cite{baroni_phonons_2001}. Within the density-density approximation, they are given by
\begin{align}
    & V_{nm}(\bR) = \langle \omega_{n\bR} \omega_{m\bzero} | v | \omega_{n\bR} \omega_{m\bzero}\rangle \nn \\
    & W_{nm}(\bR) = \langle \omega_{n\bR} \omega_{m\bzero} | \epsilon^{-1}v | \omega_{n\bR} \omega_{m\bzero} \rangle;
    \label{eq:VW_me}
\end{align}
while the bare Coulomb $V_{nm}(\mathbf{R})$ can be easily evaluated in reciprocal space taking advantage of fast Fourier transform, the screened term $W_{nm}(\mathbf{R})$ requires the knowledge of the dielectric response of the system $\epsilon^{-1}$ and represents the most time-consuming part of the calculation. The explicit evaluation of the static dielectric function $\epsilon^{-1}(\br, \br')$ can be avoided by noticing that the second line in  Eq.~\eqref{eq:VW_me} can be recast into a linear response problem describing the response of the system to an external perturbation given by the Hartree potential generated by the Wannier orbital density $\rho_{n \bzero}$:
\begin{align}
    W_{nm}(\mathbf{R}) = & \langle \rho_{n\bR}|\epsilon^{-1}v| \rho_{m\mathbf{0}}\rangle \nonumber \\ 
           = & \langle \rho_{n\bR} |v + v\chi_{\rm RPA} v | \rho_{m\mathbf{0}} \rangle \nonumber \\
           = &  \langle V_{n\bR}^{\rm H} | \rho_{m\mathbf{0}} \rangle+ \langle V_{n\bR}^{\rm H} | \chi_{\rm RPA}|V_{m\mathbf{0}}^{\rm H}\rangle \nonumber \\
           = & V_{nm}(\mathbf{R}) + \langle V_{n\bR}^{\rm H} | \Delta^{m\mathbf{0}}\rho\rangle
\end{align}
In the expression above we used the definition of the RPA dielectric matrix ${\epsilon^{-1} = 1 + v\chi_{\rm RPA}}$, where $\chi_{\rm RPA}$ is the 
irreducible density-density response function of the system;
${\Delta^{m\bzero} \rho(\br) = \int d\br' \chi_{\rm RPA}(\br, \br') V_{m\bzero}^{\rm H}(\br')}$ is by definition the density response induced in the systems due to the ``perturbing potential'' ${V_{m\mathbf{0}}^{\rm H} (\br) = \int d\br' v (\br, \br') \rho_{m\bzero}(\br')}$. This perturbation represents the Hartree potential generated by the Wannier orbital density $\rho_{m\bzero}(\mathbf{r})$. 
Similarly, $V_{n\bR}^{\rm H} (\br) = \int d\br'v (\br, \br') \rho_{n\bR}(\br')$ represent the Hartree potential generated by the Wannier orbital density $\rho_{n\bR}(\mathbf{r})$. 

Using the definition of Wannier function the generic Wannier orbital density $\rho_{n\bR}$ can be decomposed into a sum of independent monochromatic contributions:
\begin{align}
    \rho_{n\bR}(\br) & = |\omega_{n\bR}(\br)|^2  = \left| \frac{1}{N_{\bk}}\sum_{\bk} e^{-i \bk \cdot \bR} \tilde{\psi}_{n\bk}(\br)\right|^2 \nonumber \\
                            & = \frac{1}{N_{\bq}}\sum_{\bq} e^{i \bq \cdot \br} \left\{ e^{-i \bq \cdot \bR} \frac{1}{N_{\bk}}\sum_{\bk} \tilde{u}^*_{n\bk}(\br)\tilde{u}_{n\bk+\bq}(\br) \right\} \nonumber \\
                            & = \frac{1}{N_{\bq}}\sum_{\bq} e^{i \bq \cdot \br} \rho_{\bq}^{n\bR}(\br)\ ,
    \label{eq:wann_orb_dens}
\end{align}
In this expression $\tilde{u}_{n\bk}(\br)=e^{-i \bk\cdot\br} \tilde{\psi}_{n\bk}(\br)$ and $\rho_{\bq}^{n\bR}(\br) = \rho_{\bq}^{n\bR}(\br+\bR')$ are the periodic part of the Bloch wavefunction in the Wannier gauge and of the Wannier density, respectively.
The perturbing potential $V_{m\mathbf{0}}^{\rm H} (\br)$ acquires the same periodic structure as the Wannier density and can also be decomposed into a sum of monochromatic perturbations in the primitive cell, $V_{ m\mathbf{0}}^{\rm H}(\br) = \sum_{\bq} e^{i\bq \cdot \br} V_{m\mathbf{0},\bq}^{\rm H}(\br)$ with 
\begin{equation}
  V_{m\mathbf{0},\bq}^{\rm H}(\br) = \int d\br' v^{\bq}(\br, \br') \rho^{m\bzero}_{\bq}(\br') .
 \label{eq:Vpert_KC}
\end{equation}
where
\begin{equation}
    v^{\bf q}(\textbf{r},\textbf{r}') = \frac{e^2e^{-i\bq\cdot(\br-\br')}}{|\br - \br'|}
\end{equation}
The total density variation $\Delta^{m\bzero} \rho(\br')$ induced by the bare perturbation $V_{m\bzero}^{\rm H}(\br)$ reads
\begin{align}
    \Delta^{m\bzero} \rho(\br) & = \int d\br' \chi_{\rm RPA}(\br, \br') V_{m\mathbf{0}}^{\rm H}(\br') \nn \\
     & =  \int d\br' \chi_{\rm RPA}(\br, \br') \sum_{\bq} e^{i\bq \cdot \br'}  V_{m\mathbf{0},\bq}^{\rm H}(\br') \nn \\
     & = \sum_{\bq} e^{i\bq \cdot \br} \Delta^{m\bzero}_{\bq}\rho (\br)
    \label{eq:dens_var}
\end{align}
where we used the fact that for a periodic system $\chi_{\rm RPA}$ can be decomposed into monochromatic contributions $\chi_{\rm RPA}(\br ,\br') = \sum_{\bq} e^{i\bq \cdot (\br - \br')}\chi_{\rm RPA}^{\bq}(\br ,\br')$.
The density variation at a given $\bq$-point is given by:
\begin{align}
    \Delta_{\bq}^{m\bzero} \rho(\br) & = \int d\br' \chi_{\bq}(\br,\br') V_{m\mathbf{0},\bq}^{\rm H}(\br') \nn \\ 
    &= \sum_{v\bk} \psi_{v\bk}^*(\br) \Delta^{m\bzero} \psi_{v\bk+\bq}(\br) + c.c.
    \label{eq:dens_var_per}
\end{align}
where $ \Delta^{m\bzero}\psi_{v\bk+\bq}(\br)$ is the first-order variation of the KS orbital $\psi_{v\bk}(\br)$ due to the perturbation (the bare one plus, within the RPA, the SCF response in the Hartree potential), and it is given by the solution of the following linear problem~\cite{baroni_phonons_2001}:
\begin{align}
    & \left( \mathcal{H} -\varepsilon_{v\bk} \right) \Delta^{m\bzero} \psi_{v\bk+\bq}(\br) = \nn \\
     = & -\left[V_{m\mathbf{0},\bq}^{\rm H}(\br) + \Delta^{m\bzero}_{\bq} V_{\rm H}(\br)\right]\psi_{v\bk}(\br)
    \label{eq:lin_eq}
\end{align}
where $\mathcal{H}$ is the ground state KS Hamiltonian, $\varepsilon_{v\bk}$ the KS eigenvalues, and 
\begin{equation}    \Delta^{m\bzero}_{\bq} V_{\rm H}(\br) = \int d \br ' v^{\bq}(\br, \br') \Delta^{m\bzero}_{\bq}\rho (\br')
    \label{eq:Delta_Vscf}.
\end{equation} 


%

The advantage of this implementation compared to standard one that requires the explicit evaluation of the density-density response matrices~\cite{nakamura_respack_2021}, is that there is no explicit reference to empty states and no need to explicitly compute, store and manipulate large response matrices. 
This has been implemented in a dedicated version of the \verb|KCW| package~\cite{gitlab_KCW_Wme} of the \QE\,distribution.

\begin{figure*}[t]
  \centering
  \includegraphics[width=0.30\textwidth]{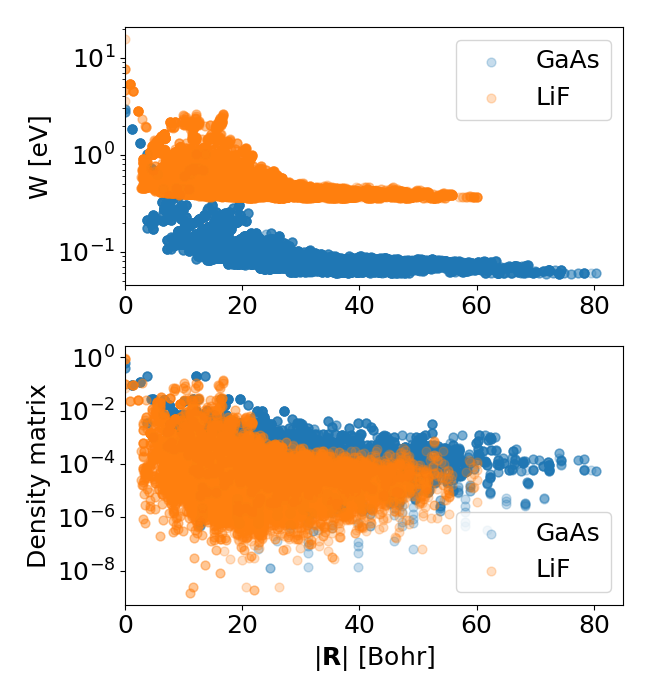}
  \includegraphics[width=0.30\textwidth]{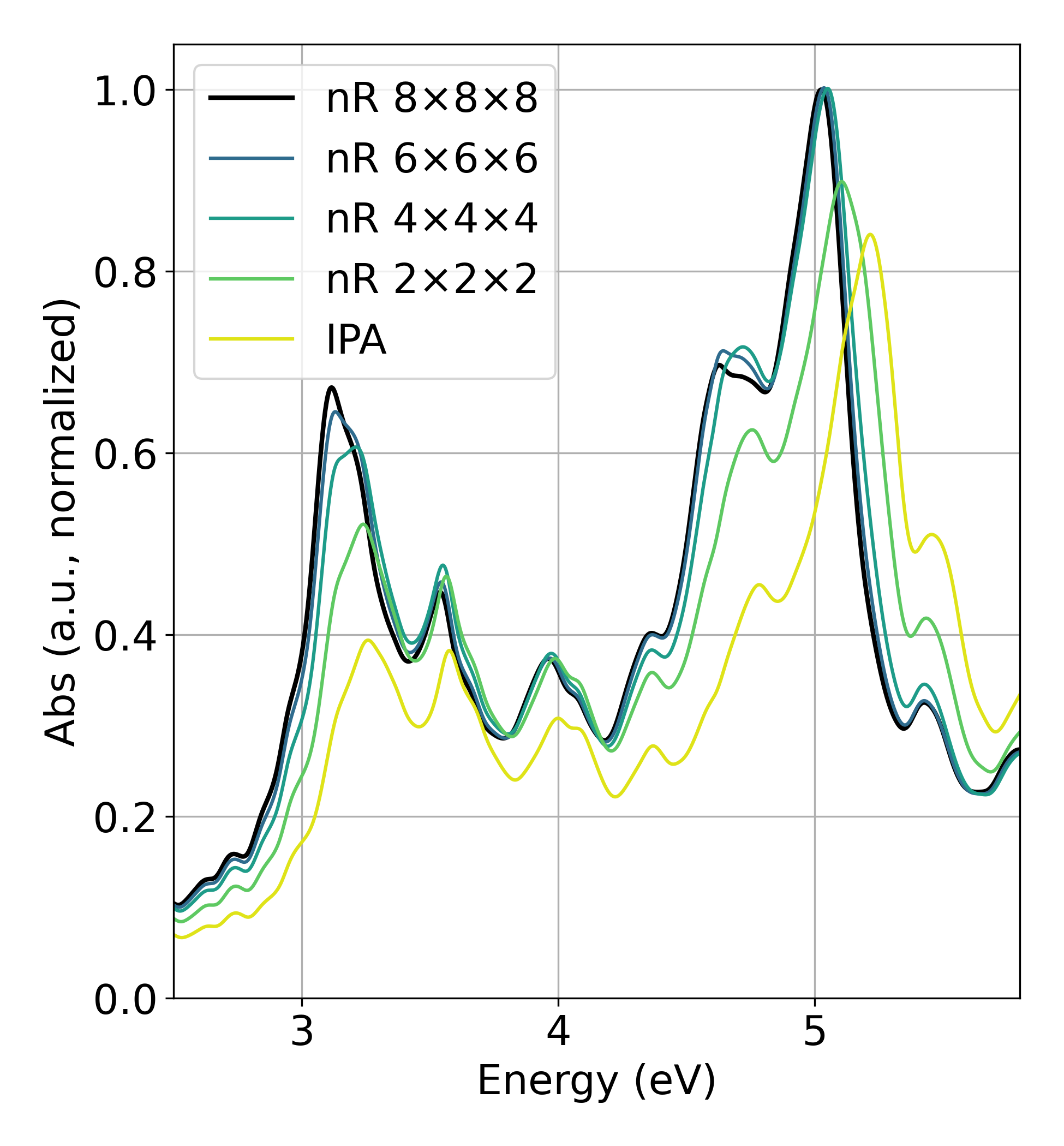}
  \includegraphics[width=0.30\textwidth]{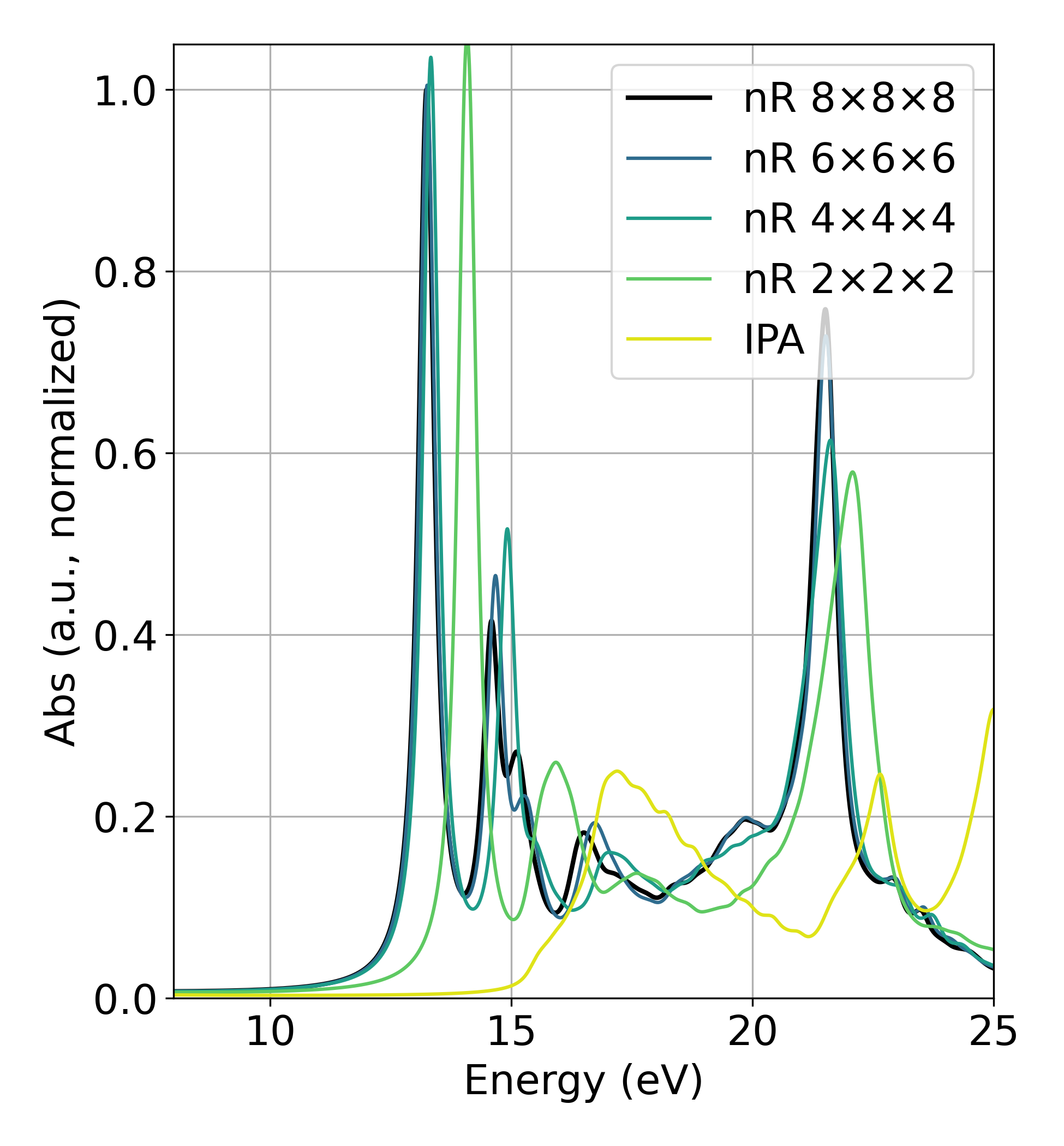}
  \caption{Left panel: decay of the density-density Coulomb integrals $W_{nm}(\bR)$ (top panel) and of the equilibrium density matrix $\rho_{nm}(\bR, t=0)$ (bottom panel) for GaAs and LiF as a function of the Wannier center distance ($\mathbf{R}$) for all the Wannier functions in our models. Middle panel: convergence of the absorption spectrum of GaAs at the HSEX level of approximation cutting-off the Coulomb interaction at increasingly large $\mathbf{R}$ shells. Right panel: convergence of the absorption spectrum of LiF at the HSEX level of approximation cutting-off the Coulomb interaction at increasingly large $\mathbf{R}$ shells. $n\bR\,N_1{\times}N_2{\times}N_3$ denotes the real-space $\mathbf{R}$-grid size at which the Coulomb-interaction is cut off.}
  \label{fig:GaAs-LiF_WofR}
\end{figure*}

\section{Results and discussion}
We present in this section the linear and non-linear responses of few paradigmatic test cases, highlighting the advantages and limitations of our approach. In particular we analyze the absorption spectrum of two standard semiconductors, namely silicon (Si) and gallium arsenide (GaAs), and that of the wide band-gap insulator lithium fluoride (LiF). We then look at responses beyond the linear regime and in particular we investigate high-harmonic generation in silicon and lithium fluoride and how this is affected by excitonic effects.

\subsection{Numerical details}

All {\it ab-initio} calculations are performed using the plane-wave (PW) and pseudopotential (PP) method as implemented in the \QE{} package~\cite{giannozzi_quantum_2009, giannozzi_advanced_2017}. The LDA functional is used as the underlying density-functional approximation for all the qKI calculations. LDA scalar relativistic Optimized Norm-conserving Vanderbilt PPs~\cite{hamann_optimized_2013,hamann_erratum_2017} from the DOJO library~\cite{van_setten_pseudodojo_2018} are used to model the interaction between the valence electrons and the nucleus plus the core electrons~\footnote{The LDA pseudopotentials are available at \href{http://www.pseudo-dojo.org/}{www.pseudo-dojo.org}. Version 0.4.1., standard accuracy}. For all the systems we used the experimental lattice parameters: $a=5.44$ \AA, $a=5.65$ \AA, and $a=4.03$ \AA {} for Si, GaAs, and LiF, respectively.
Koopmans band structure calculations are performed using the \texttt{Koopmans} package~\cite{linscott_koopmans_2023}. 
For the DFT, qKI, and HSEX kernel calculations a $8 \times 8 \times 8$ $\bk$-point mesh is used for the sampling of the Brillouin zone. The kinetic energy cutoff for the PW expansion of the wave-functions is set to $E_{\rm cut} = 60$ Ry (240 Ry for the density and potentials expansion) for Si and to $E_{\rm cut} = 80$ Ry (320 Ry for the density and potentials expansion) for GaAs and LiF. 
The long-ranged nature of the Coulomb kernel leads to a slow convergence of the Coulomb matrix elements as a function of the BZ sampling. To speed-up the convergence we adopt the approach devised by Gygi and Baldereschi (GB)~\cite{gygi_quasiparticle_1989} generalized to dielectric environment.~\cite{colonna_jctc_2022, rurali_theory_2009}

Maximally localized Wannier functions and related properties are computed using the Wannier90 code~\cite{pizzi_wannier90_2020}. For absorption calculations, a standard 8-band model featuring $sp3$-like MLWFs and spanning the valence manifold and the low-energy region of the conduction manifold is used for both Silicon and GaAs, while for LiF a 10-MLWF model spanning the entire occupied manifold (within the pseudo-potentials approximation) and the low-energy region of the conduction manifold is used. For the HHG calculations, different reduced Wannier models are employed for the two materials. For LiF, we use a 7-band model consisting of three $p$-like Wannier functions centered on the F atom and one $s$-like plus three $p$-like Wannier functions centered on the Li atom. For Si, we still use an 8-band model but now featuring atom-centered $s$- and $p$-like Wannier functions. These reduced models differ from the one used for absorption calculation because HHG simulations are particularly sensitive to the details of the electronic structure representation. In particular, although the underlying crystals are centrosymmetric, the construction of MLWFs does not necessarily preserve all crystal symmetries exactly, as the localization procedure can introduce small symmetry-breaking components into the Wannier representation. These residual symmetry violations can artificially break inversion symmetry, leading to spurious even-order harmonics in the calculated HHG spectra, despite such harmonics being forbidden in centrosymmetric materials like Si and LiF. To eliminate these artifacts, the quasi-particle Hamiltonian and the position operator in the Wannier basis are symmetrized using the WannierBerri code.\cite{tsirkin_high_2021} Since the symmetrization procedure implemented in WannierBerri requires Wannier functions to be atomic-like and centered on atomic sites, the reduced Wannier models described above are used for the HHG calculations. Enforcing the proper symmetries on the quasi-particle Hamiltonian and position operator removes the numerical artifacts discussed above and ensures that the calculated HHG response is consistent with the symmetry of the crystal (see Supporting Information).


For the real-time part, we propagate the density matrix in time using Runge-Kutta 4th order and Adams-Bashforth 5th order algorithm. Thanks to Wannier interpolation, we are able to use denser grids in $\bf k$ space, under which we tested convergence (see Supporting Informations). For linear responses, a sparse grid as the ones used in DFT simulations is typically enough, while as the intensity of the external field increases, denser grids are needed and Wannier interpolation becomes crucial. In practice, we used a $24^3$ $\bk$-point grid for the linear response calculations and a $32^3$ $\bk$-point grid for the non-linear one.


\begin{figure*}[t]
  \centering
  \includegraphics[width=0.98\textwidth]{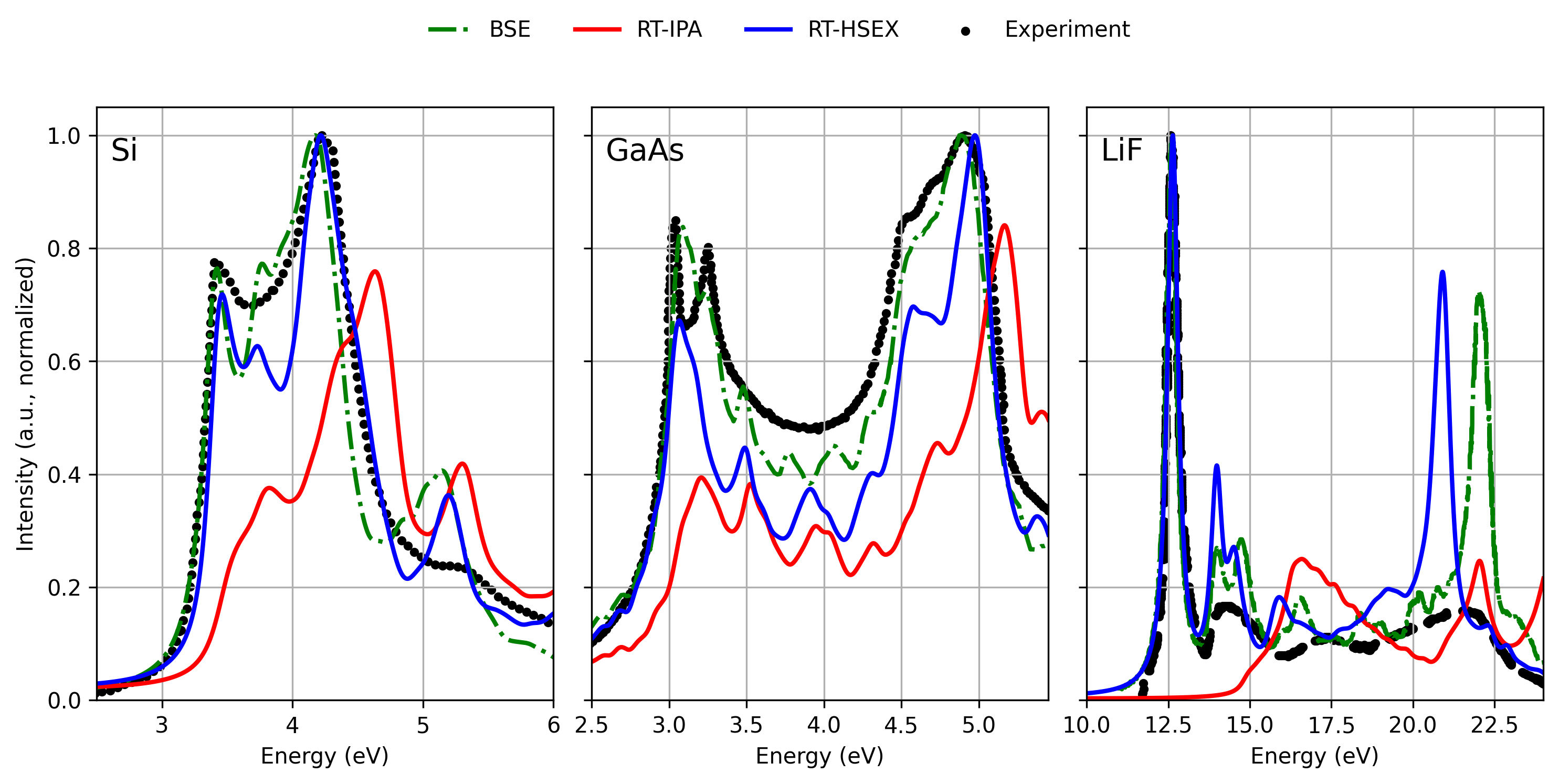}
  \caption{Absorption spectrum of the studied systems (Silicon, left panel; GaAs middle panel; LiF right panel) computed at different level of theory: IPA (red solid line), HSEX (blue solid line), and BSE~\cite{dong_machine_2021, alliati_double_2022, benedict_optical_1998} (green dash-dotted line). Experimental results (black dots) from Ref.~\cite{aspnes_dielectric_1983, lautenschlager_interband_1987, roessler_optical_1967}. All the theoretical spectra have been aligned to the first experimental peak.}
  \label{fig:ABS}
\end{figure*}

\subsection{Short range nature of the self-energy}
The left panel of Fig.~\ref{fig:GaAs-LiF_WofR} shows how the density-density screened Coulomb interaction $W_{nm}(\mathbf{R})$ and the density matrix decay as a function of the Wannier distance ($\mathbf{R}$) for GaAs (small band gap semiconductor) and LiF (wide band gap insulator). For GaAs the screened Coulomb interaction decays quickly to zero while the density matrix is relatively long range; on the other side for LiF the opposite is true with $W$ being more long range and the density matrix showing a faster decay compare to the case of GaAs. This suggests that the HSEX self-energy in Eq.~\eqref{eq:HSEX_selfenergy} is in both cases relatively short range as either the $W$ (in the case of metals or small band gap insulators) or the density matrix (in the case of large band gap materials) are short range dominating the decay of the HSEX self-energy. This is confirmed in the middle and right panels of Fig.~\ref{fig:GaAs-LiF_WofR} where we show the evolution of the absorption spectra of GaAs and LiF, respectively, when larger and larger contribution to the $V(\mathbf{R})$ (and $W(\mathbf{R})$) matrices are added in real space, going from the IPA case (no coulomb interaction at all) to the upper limit of a $8\times 8 \times 8$ \textit{ab-initio} grid. For the case of LiF cutting-off the Coulomb interaction beyond a $4\times 4\times 4$ grid in real space (corresponding to a ~30 Bohr cut-off radius) leads to an absorption spectrum that is almost indistinguishable from the one computed with full matrices on a $8\times 8\times 8$ $\mathbf{R}$-grid (corresponding to a ~60 Bohr cut-off radius) despite the  Coulomb interaction is far from being close to zero.


\begin{figure*}[h]
  \centering
  \includegraphics[width=0.95\textwidth]{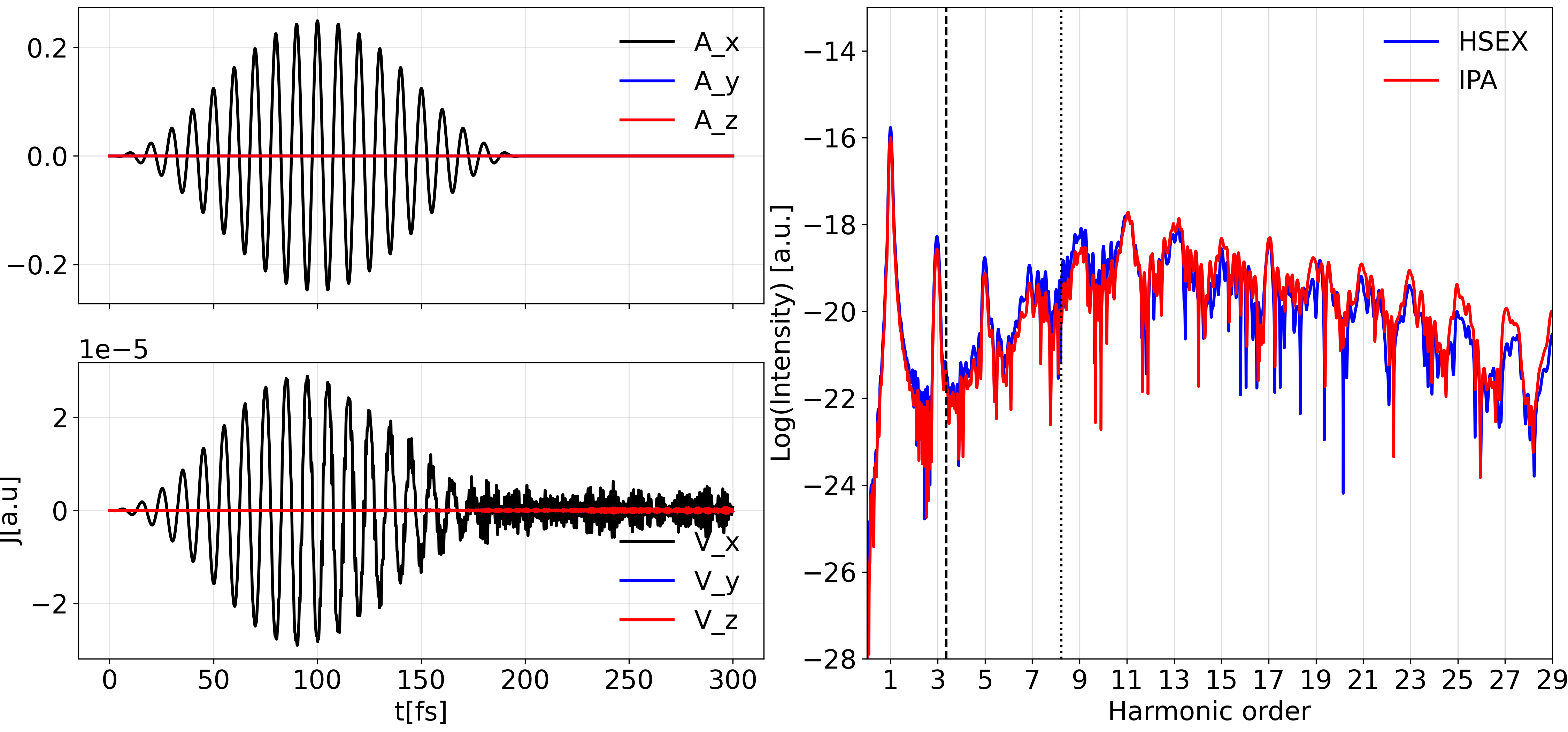}
  \caption{The applied vector potential (left top) of a pulse with a full duration of 200 fs at $\lambda = 3000$ nm and peak intensity $5 \times 10^{11}$ W/cm$^2$, and the induced current (left, bottom) in Si at the HSEX level of approximation. The HHG spectra for Si (right) at different level of approximations: IPA (solid red) and HSEX (solid blue). Dashed vertical lines mark the indirect and direct quasi-particle qKI band gaps.}
  \label{fig:Si_HHG}
\end{figure*}

\subsection{Linear regime}
\label{sec:Linear_regime}

We begin by benchmarking our approach in the optical linear response, which is well established for the above mentioned materials. Within the HSEX approximation, and in the limit of weak external electric fields, Eq.~\eqref{eq:EOM} reduces to the Bethe-Salpeter equation~\cite{attaccalite_real-time_2011}. Fig.~\ref{fig:ABS} shows the absorption spectra for all the systems investigated in this work calculated at the IPA and HSEX level of approximations (within the density-density approximation for the HSEX kernel (Eq.\eqref{KC::4})) using the method presented in this work and a comparison with reference G$_0$W$_0$+BSE~\cite{dong_machine_2021, alliati_double_2022, benedict_optical_1998} and experimental results~\cite{aspnes_dielectric_1983, lautenschlager_interband_1987, roessler_optical_1967}. In the comparison, we aligned the computed spectra with the experimental ones by matching the onset of the first peak (see Supporting Information for the un-shifted spectra), and we normalized to one the highest peak of the BSE, exp, and HSEX spectra). As expected, the IPA spectrum is completely missing the excitonic peak; adding HSEX correlations a better agreement with reference many-body calculations and experimental results is achieved. The residual discrepancy between HSEX and G$_0$W$_0$+BSE mainly amounts to a different relative intensity of the peaks and it is due to two effects: first, the underlying quasi-particle Hamiltonian comes from two different {\it ab-initio} approaches (qKI vs GW); second, the density-density approximation (Eq.~\eqref{KC::4}) our method relies on implicitly requires the Wannier functions to be perfectly localized and as little overlapping as possible, condition that might not be perfectly met. Nevertheless, we clearly observe the appearance of the excitonic peaks indicating that the density-density approximation is already accounting to a satisfactory extent for the correct screened electron-hole interaction. This effect is particularly relevant for large band-gap insulating materials where the absorption spectrum at low energy is dominated by excitons (bound electron-hole pairs) appearing inside the quasiparticle band gap. This is clearly observed from our HSEX simulations of LiF in  the right panel of Fig.~\ref{fig:ABS}. Compared to the rt-IPA simulation the HSEX shows a strong resonance at ~13.3 eV (see Fig.~\ref{fig:GaAs-LiF_WofR}, right panel and Supporting Information for the unshifted absorption spectra) well below the quasi-particle qKI band gap (15.3 eV). 


\subsection{Non-linear regime}

Real-time approaches enable the calculation of the optical response of materials beyond the linear regime. Ultrafast processes constitute a specific class of such phenomena, where the optical response of the probe pulse is modified by a prior interaction of the pump pulse with the system. In this case, the measured signal involves, at least, a second-order optical process. However, real-time approaches are not limited to weak or perturbative regimes, they also allow one to describe systems driven by strong laser fields, where the response becomes intrinsically nonperturbative and cannot be captured within a conventional multiphoton expansion. High-harmonic generation is a prominent nonlinear optical phenomenon arising from the strong-field interaction between light and matter, in which an intense driving pulse induces the emission of radiation at integer multiples of its fundamental frequency \cite{amini2019}. The computation of HHG emission is highly sensitive to the underlying induced dynamics, making it an ideal benchmark to assess the capabilities of our approach. The strong-field regime involves the participation of multiple electronic bands, as well as pronounced intraband dynamics, which in turn requires a dense sampling of the reciprocal space for accurate convergence.

\begin{figure*}[t]
  \centering
  \includegraphics[width=0.95\textwidth]{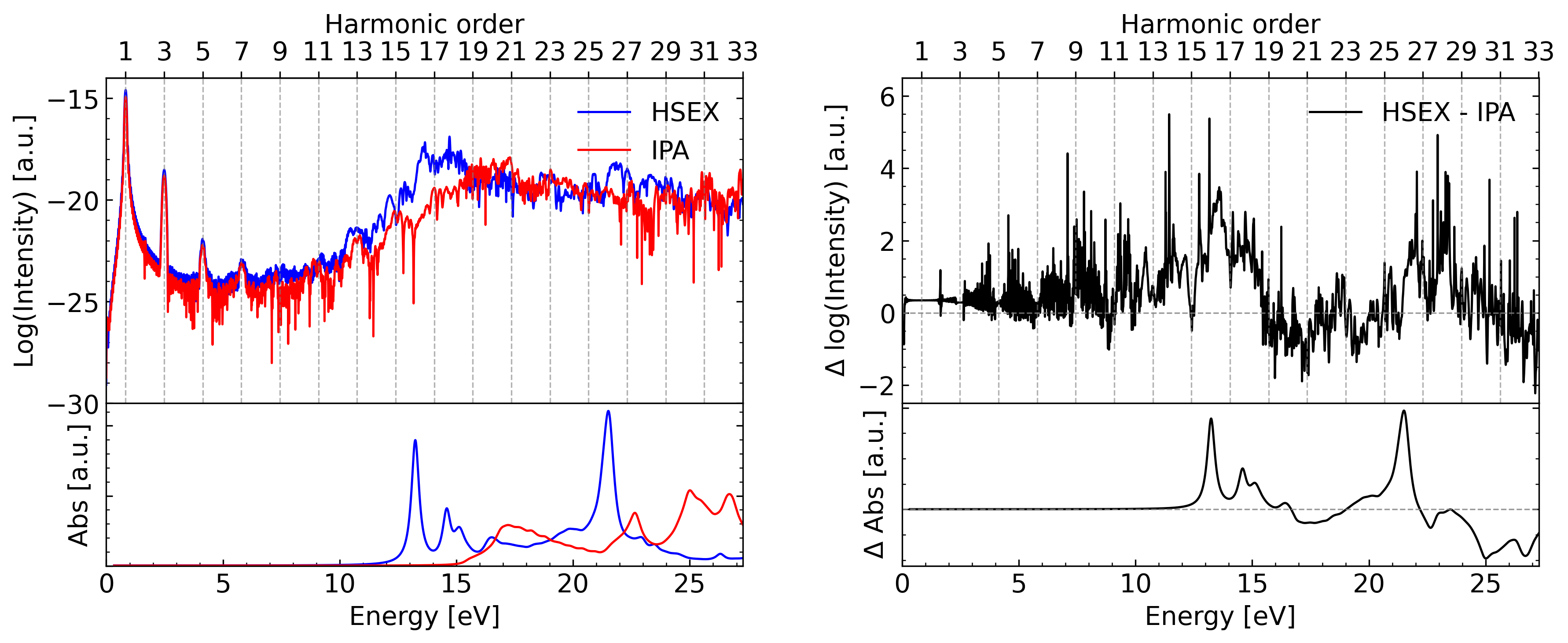}
  \caption{Left panel: HHG spectrum of LiF computed at different level of theory: IPA (red solid line), HSEX (blue solid line). The inset at the bottom shows the corresponding absorption spectra. Right panel: Difference between the HSEX and IPA HHG spectra. The inset at the bottom show the corresponding difference in the absorption spectra.}
  \label{fig:LiF_HHG}
\end{figure*}

High-harmonic generation was first studied in atomic systems and was later extended to solid-state materials in the early 2000s \cite{vampa2017}. Within the independent-particle approximation, the microscopic mechanism is commonly interpreted in terms of interband polarization and intraband acceleration of Bloch electrons~\cite{amini2019}.
In this picture, electrons are promoted from valence to conduction bands through strong-field excitation, accelerated by the external field, and eventually recombine with the holes left behind, emitting radiation at integer multiples of the driving frequency. As a consequence, several characteristics of the HHG spectrum can be directly related to the underlying electronic band structure, and the harmonic emission can often be understood within a semiclassical extension of the three-step model to periodic solids~\cite{amini2019,tancogne-dejean_impact_2017}. Although HHG emission is typically well understood within the IPA, the role of excitonic effects has also been recently investigated~\cite{attaccalite_nonlinear_2013, gruning_second_2014, attaccalite_excitonic_2017, chang_lee_many-body_2024, attaccalite_second-harmonic_2019, hou_data-driven_2025, jensen_high-harmonic_2024}. A key feature is the enhancement of the emission near photon energies corresponding to excitonic resonances \cite{molinero2024}. To the best of our knowledge, RT-HSEX simulations have so far been mainly limited to two-dimensional materials \cite{chang_lee_many-body_2024}. However, achieving a fully \textit{ab-initio} description of HHG spectra that accurately incorporates excitonic effects in three-dimensional systems remains computationally demanding and constitutes a significant challenge. Addressing this limitation is essential for a quantitative interpretation of experiments, which have been predominantly performed in bulk materials.\\
The calculated HHG spectrum of silicon in the presence of an intense near infrared laser pulse is shown in Fig.~\ref{fig:Si_HHG}, for both the IPA and HSEX lvele of approximation. 
We observe harmonic emission below the band gap, with the intensity decreasing as the harmonic order increases, as expected in the multiphoton regime. Above the band gap, however, the harmonic yield no longer follows this trend, consistent with the onset of the strong-field regime. Comparison between the IPA and HSEX results shows that there is very little difference in the spectra. For conventional semiconductors such as Si, where screening is efficient and excitonic effects are relatively weak, the IPA picture supplemented by quasiparticle corrections provides a reasonable description of the nonlinear response. 
Real-time TDDFT simulations within common local or semi-local density functionals have successfully reproduced the harmonic emission in Si over a broad range of driving wavelengths, supporting the validity of this quasiparticle picture ~\cite{suthar_role_2022, tancogne-dejean_impact_2017, freeman_high-order_2022}. In conclusion, the main role of many-body interactions is to renormalize the quasiparticle band structure and transition energies, while the overall characteristics of the HHG spectra remain largely determined by the underlying bands \cite{tancogne-dejean_impact_2017}.

In wide-gap insulators and weakly screened materials, however, the Coulomb attraction between excited electrons and holes gives rise to strongly bound excitons, indicating that the elementary optical excitations are no longer adequately described as independent electron-hole pairs. 
Recent studies in two-dimensional materials have shown that electron-hole interactions can substantially enhance and reshape HHG spectra \cite{molinero2024,chang_lee_many-body_2024}. As we demonstrate in the following, similar effects can also arise in three-dimensional materials characterized by strong electron-hole interactions.

Lithium fluoride provides a prototypical example of such three-dimensional counterpart. Fig.~\ref{fig:LiF_HHG} shows the simulated HHG spectrum of LiF (top panel) computed starting from the qKI Hamiltonian, and using the IPA and the HSEX approximation for the self-energy. The quasi-particle qKI band gap energy is 15.3 eV, and we compute the strong-field optical response of LiF induced by an external pulse with a full duration of 100 fs at $\lambda = 1500$ nm and a peak intensity of $1.2 \times 10^{13}$ W/cm$^2$. 

At variance with Silicon, the IPA and HSEX spectra show significant differences especially in the energy range of the excitonic peaks. As shown in Fig.~\ref{fig:ABS} (right panel) and in the inset of the left panel of Fig.~\ref{fig:LiF_HHG}, the optical absorption spectrum of LiF is dominated by strongly bound excitons.
The same strong electron-hole interaction responsible for the excitonic resonances also affects the nonlinear dynamics and the HHG process. 
In the right panel 
of Fig.~\ref{fig:LiF_HHG} one can see that the difference between the HSEX and IPA HHG spectra correlates significantly with the same difference in the absorption spectra. The HSEX HHG signal is strongly enhanced in correspondence of the excitonic resonances (in particular around 13.3 eV and 21.5 eV corresponding to the region around the 17$^{\rm th}$ and 27$^{\rm th}$ harmonics). This results shows that a consistent description of HHG in LiF and, more  generally, in systems with weak screening effects requires an explicit treatment of electron-hole correlations, and that the interpretation of HHG solely in terms of quasiparticle band structures becomes questionable. 


\section{Conclusions}
We have presented an efficient, fully \textit{ab-initio} framework for simulating nonequilibrium electron dynamics and ultrafast optical spectroscopy in extended systems. The approach combines Koopmans-compliant spectral functionals, with real-time propagation of the one-electron density matrix in a Wannier basis within the Hartree plus screened exchange approximation. We introduced an efficient scheme for evaluating the bare and screened Coulomb matrix elements in the Wannier basis, the principal computational bottleneck of the method. Exploiting the spatial localization of Wannier functions and computing the screened Coulomb interaction via DFPT, substantially reduces both computational cost and memory requirements compared to standard implementations. 

We validated the method in the linear regime by computing the optical absorption spectra of silicon, gallium arsenide, and lithium fluoride, obtaining good agreement with experiment and state-of-the-art GW-BSE calculations. In all cases, HSEX significantly improves upon the independent-particle approximation, and for LiF it correctly reproduces the strongly bound excitonic peaks below the quasiparticle gap. In the nonlinear regime, we investigated high-harmonic generation (HHG) in silicon and LiF driven by intense infrared pulses. For silicon, where excitonic effects are weak, IPA and HSEX produce similar spectra, indicating that many-body interactions mainly renormalize the quasiparticle band structure. In contrast, for LiF the HSEX spectrum exhibits a clear enhancement of harmonic emission at energies corresponding to excitonic resonances, demonstrating that strong electron-hole correlations play a crucial role in the HHG response.

Overall, these results establish \verb|KCW-EDUS| as a reliable and computationally efficient route to fully \textit{ab-initio} simulations of ultrafast and strong-field-driven phenomena in solids, approaching the accuracy of real-time GW methods at significantly lower complexity. 

\suppinfo
{
The Supporting Information is available free of charge on the ACS Publications website.
Additional benchmarks complementing the validation presented in the main text, including: unshifted linear-response absorption spectra of Si, GaAs, and LiF at the   IPA, HSEX, and BSE levels compared against experiment (Figure S1); a comparison between Koopmans-compliant (qKI) and G0W0 quasiparticle band gaps (Table S1);   validation of the real-time propagation engine against reference TD-DFT high-harmonic-generation (HHG) spectra of Si (Figure S2); the effect of Wannier-Hamiltonian   symmetrization on the HHG spectra of Si and LiF (Figure S3); a comparison of absorption and HHG spectra of Si obtained with two different Wannier models (Figure   S4); and convergence tests of the simulated HHG spectra of Si and LiF with respect to the k-point sampling and the real-time integration time step (Figures S5–S8) 
}

\section*{Acknowledgments}
The authors acknowledge useful discussions with G. Stefanucci, E. Perfetto, and S. Tsirkin. 
G.C. acknowledges funding from SNSF (Swiss National Science Foundation), Mobility grant No. P500-2\_239103. D.S. acknowledges funding from the PRIN project EXATTO funded by Italian Miur (Grant No. 2022PX279E), from the MaX project co-funded by the European High Performance Computing joint Undertaking and participating countries (Grant Agreement No. 101093374), and from the TIMES project funded by the European Union’s Horizon Europe research and innovation programme under the Marie Sklodowska-Curie (Grant Agreement No. 101118915).
A.P. acknowledges the Spanish Ministry of Science, Innovation and Universities \& the State Research Agency through grants refs. PID2024-157663NB-I00 (MCIU/AEI/FEDER, UE), and the ``Severo Ochoa" Programme for Units of Excellence in R\&D (CEX2024-001445-S), and computer resources and assistance provided by the Centro de Computaci\'on Cient\'ifica (CCC-UAM) and the Red Espa\~nola de Supercomputaci\'on (RES) under projects refs. FI-2026-1-0038, FI-2025-2-0036, FI-2024-3-0011, and FI-2024-2-0034. 
N.C. acknowledges partial support from the NCCR MARVEL, a National Centre of Competence in Research, funded by the Swiss National Science Foundation (Grant No. 205602) and the Swiss National Supercomputing Centre (CSCS) for high-performance computing resources under the CSCS-PSI agreement.

\section*{Data Availability}
All the input files and data used to produce the results and figures of this work are available at the Materials Cloud Archive.~\cite{MC_archive}

\newpage 
\bibliography{biblio}
\newpage

\end{document}